\documentclass[conference,a4paper]{IEEEtran}
\IEEEoverridecommandlockouts
\usepackage{cite}
\usepackage{amsmath,amssymb,amsfonts}
\usepackage{algorithmic}
\usepackage{graphicx}
\usepackage{textcomp}
\usepackage{xcolor}
\usepackage[a4paper, total={184mm,239mm}]{geometry}
\def\BibTeX{{\rm B\kern-.05em{\sc i\kern-.025em b}\kern-.08em
    T\kern-.1667em\lower.7ex\hbox{E}\kern-.125emX}}
\begin{document}
\bstctlcite{IEEEexample:BSTcontrol}

\makeatletter
\def\endthebibliography{%
  \def\@noitemerr{\@latex@warning{Empty `thebibliography' environment}}%
  \endlist
}
\makeatother

\title{Security layers and related services within the Horizon Europe NEUROPULS project\\
\thanks{This work has received funding from the European Union’s Horizon Europe research and innovation program under grant agreement No. 101070238. Views and opinions expressed are however those of the author(s) only and do not necessarily reflect those of the European Union. Neither the European Union nor the granting authority can be held responsible for them. It was also supported by the ANR within the PHASEPUF Project
ANR-20-CE39-0004.}
}
\author{
\IEEEauthorblockN{Fabio~Pavanello$^1$, Cedric~Marchand$^2$, Paul~Jimenez$^2$, Xavier~Letartre$^2$, Ricardo~Chaves$^3$, Niccolò~Marastoni$^{4}$, \\ Alberto~Lovato$^{4}$, Mariano~Ceccato$^{4}$,
George~Papadimitriou$^{5}$, Vasileios~Karakostas$^{5}$, Dimitris~Gizopoulos$^{5}$,\\
Roberta~Bardini$^{6}$, Tzamn~Melendez~Carmona$^{6}$, Stefano~Di~Carlo$^{6}$, Alessandro~Savino$^{6}$, Laurence~Lerch$^7$,\\ Ulrich~Ruhrmair$^7$, Sergio~Vinagrero~Gutiérrez$^8$, Giorgio~Di~Natale$^8$, Elena~Ioana~Vatajelu$^8$}\\
\IEEEauthorblockA{
$^1$Univ. Grenoble Alpes, Univ. Savoie Mont Blanc, CNRS, Grenoble INP, IMEP-LAHC, Grenoble, France\\
$^2$Univ. Lyon, Ecole Centrale de Lyon, INSA Lyon, Université Claude Bernard Lyon 1, CPE Lyon, CNRS, INL, Ecully, France\\
$^3$INESC-ID,IST, ULisboa, Lisbon, Portugal\\
$^4$Department of Computer Science, University of Verona, Verona, Italy\\
$^5$Department of Informatics and Telecommunications, National and Kapodistrian University of Athens, Athens, Greece\\
$^6$Politecnico di Torino, Control and Computer Eng. Department, Italy\\
$^7$Technical University of Berlin, Berlin, Germany\\
$^8$Univ. Grenoble Alpes, CNRS, Grenoble INP, TIMA, 38000 Grenoble, France\\
}
}
\maketitle

\begin{abstract}
In the contemporary security landscape, the incorporation of photonics has emerged as a transformative force, unlocking a spectrum of possibilities to enhance the resilience and effectiveness of security primitives. This integration represents more than a mere technological augmentation; it signifies a paradigm shift towards innovative approaches capable of delivering security primitives with key properties for low-power systems. This not only augments the robustness of security frameworks, but also paves the way for novel strategies that adapt to the evolving challenges of the digital age.

This paper discusses the security layers and related services that will be developed, modeled, and evaluated within the Horizon Europe NEUROPULS project. These layers will exploit novel implementations for security primitives based on physical unclonable functions (PUFs) using integrated photonics technology. Their objective is to provide a series of services to support the secure operation of a neuromorphic photonic accelerator for edge computing applications.
\end{abstract}


\section{Introduction}
The growing intelligence of the current environments is closely related to the deployment and use of neural networks (NN). However, many edge computing devices, such as those related to the Internet of Things (IoT), present serious challenges in terms of computing and security~\cite{roman2018mobile,xiao2019edge}.
On the one hand, edge computing devices cannot rely on the same level of resources as more sophisticated high-end computing solutions, e.g., located in data centers, for costs and weight reasons. On the other hand, edge computing devices present multiple access points, for example, through other computing nodes in the same network, which can be exploited to carry out various types of attacks \cite{ranaweera2021survey}.
Therefore, it is crucial to develop novel security layers compatible with edge computing device constraints in terms of lightweight character, low cost, low power, and robustness against attacks. 

To address these requirements, hardware security primitives can be regarded as one of the best candidates to relax some of the constraints of classical systems, for example, where a secret key is required to be stored directly in a non-volatile memory \cite{butun2020hardware}.
Such security primitives can be used efficiently against attacks that aim to access specific memory sectors, as in the case of Spectre and Meltdown hardware vulnerabilities \cite{kocher_spectre_2019,lipp2018meltdown}. 

A well-known class of hardware primitives that offers a solution to such attacks is one of physically unclonable functions (PUFs). Originally pioneered at MIT in 2002 by two independent groups in the optical and electronic domains, these primitives can be used in conjunction with well-known security protocols to establish low-power and robust security layers \cite{pappu_physical_2002,gassend_silicon_nodate}. Although various technologies have been investigated to enable such primitives, CMOS-based PUFs have become the most widely used solution to enable a series of security services such as cryptographic key generation, secure authentication, or root-of-trust features \cite{shamsoshoara2020survey}. However, electronic solutions such as SRAM or Arbiter PUF have limitations in terms of reliability with respect to aging and fluctuations, as well as a limited number of degrees of freedom and related achievable complexity, which results in solutions prone to machine learning attacks \cite{pavanello_recent_2021}.

\begin{figure}
    \centering
    \includegraphics[width=1\linewidth]{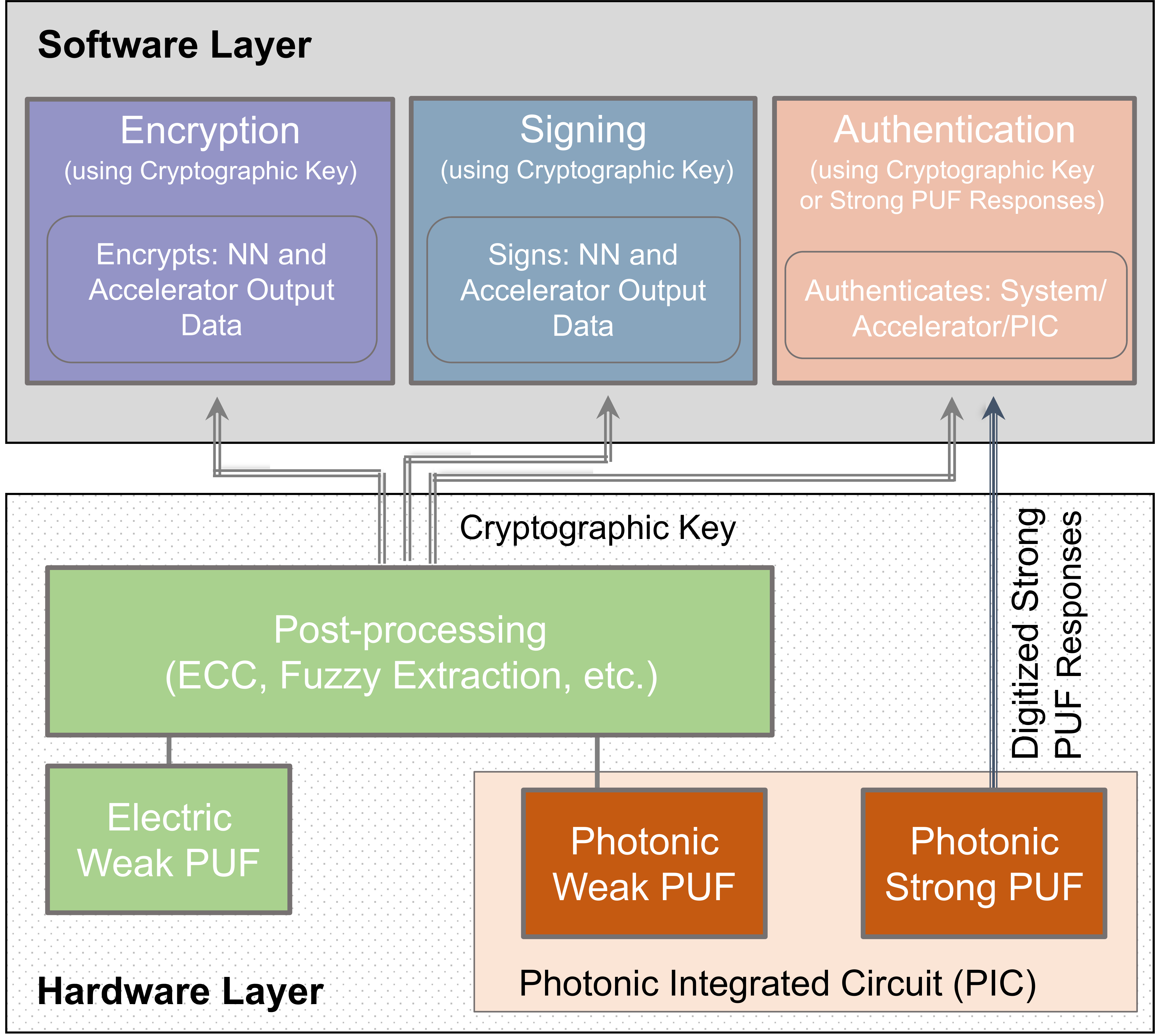}
    \caption{Hardware-Software communication flow for security services in NEUROPULS. Weak and strong PUFs target different security services and supporting circuitry.}
    \label{fig:security_HW_SW}
\end{figure}

In the Horizon Europe NEUROPULS project (which started in January 2023) \cite{10173974}, our objective was to develop security layers based on novel security primitives by leveraging the technology available in our platform. Specifically, our goal is to develop security layers based on a photonic integrated circuit (PIC). Such technology will be used to develop a neuromorphic photonic accelerator which we aim to protect using the very same photonic technology as the accelerator.
A high-level description of the security workflow in NEUROPULS is represented in Fig.~\ref{fig:security_HW_SW} where primitives based on PUFs, implemented in a PIC alongside the accelerator and in an ASIC (based on SRAM) to guarantee unique binding between the chips, provide their output to the software layer. The software layer will implement specific protocols, discussed in the following, which will provide a series of services for the secure operation of the accelerator.

In the remainder of this document, we will first discuss security primitives based on photonic technologies and their operation in Section ~\ref{sec:security_primitives}. Then, in Section~\ref{sec:services}, we will introduce the security services provided to the accelerator based on such primitives. Finally, we will qualitatively analyze the strength against attacks of the proposed security layers in Section~\ref{sec:attacks} and their modeling requirements in Section~\ref{sec:benchmarking} to precisely predict their impact on the overall performance of the system.

\section{Security Primitives and Related Metrics} \label{sec:security_primitives}
\subsection{PUFs}
Although the majority of PUF implementations are based on CMOS technology because of its native compatibility with CMOS interfaces, they present a series of limitations, as mentioned above, which stem from their digital nature, but also technology. Photonics can allow the building of PUFs that present a much larger degree of freedom (e.g., phase, polarization, amplitude) and where the information is manipulated completely differently from a classical CMOS-based PUF. Furthermore, signals are propagated in the analog domain and therefore can carry a much higher entropy than digital PUFs \cite{pavanello_recent_2021}.

\begin{figure}
    \centering
    \includegraphics[width=0.8\linewidth]{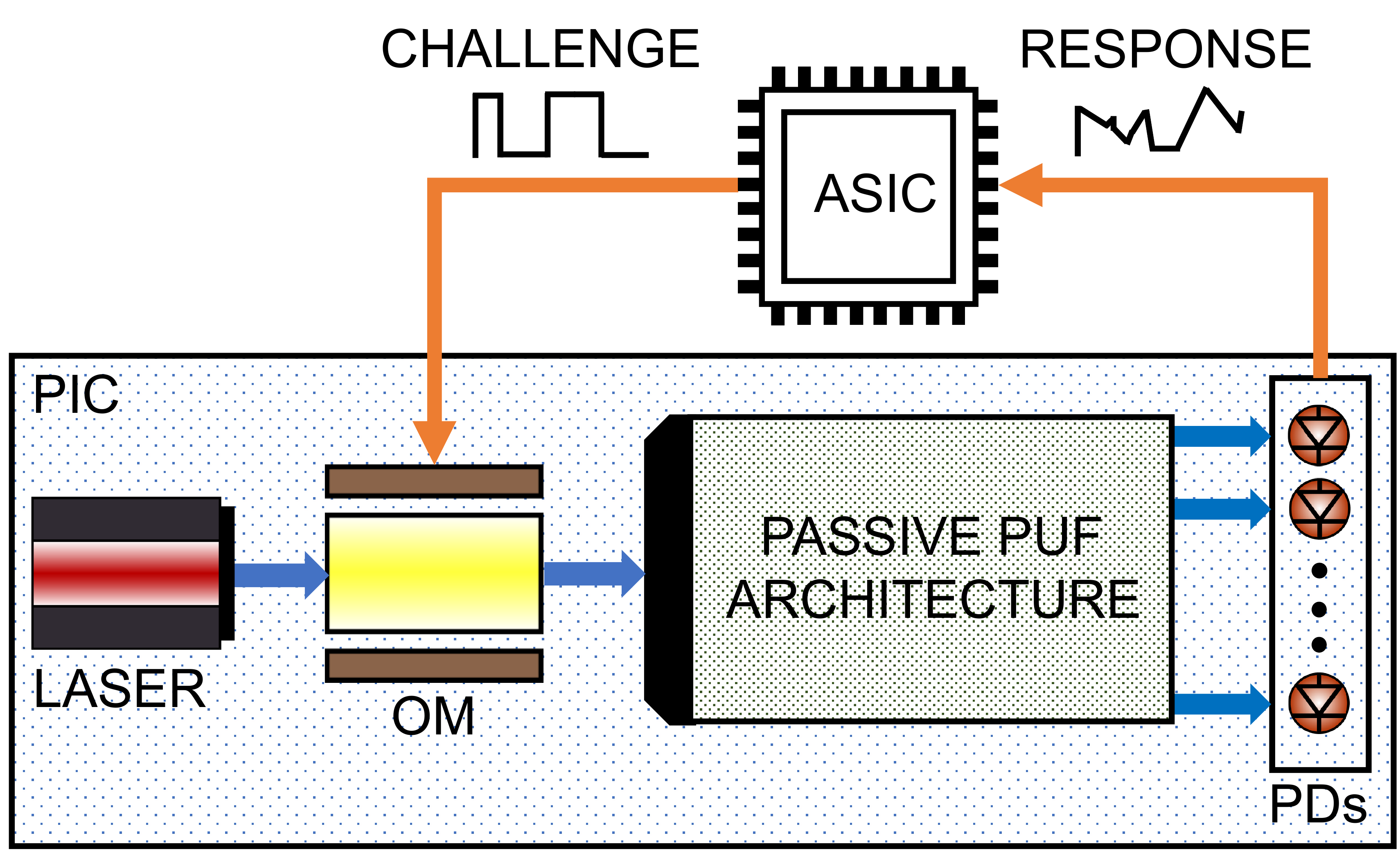}
    \caption{Schematic of the PUF operation considered in NEUROPULS. OM: optical modulator, PDs: photodiodes. The passive PUF architecture section separates the initial light beam in several different paths and scrambles them before the output. No active devices are present. In the final demonstrator, all the PIC components shown and ASIC-related PUF circuitry will constitute the actual PUF.}
    \label{fig:PUF_high_level}
\end{figure}

In NEUROPULS, we investigate various types of photonic architectures for weak and strong PUFs that can be schematically summarized in Fig.~\ref{fig:PUF_high_level}. Here, we have a telecom laser source that is modulated by means of an optical modulator (OM) driven by an ASIC. The light beam enters a passive architecture (no active devices are present) featuring a large number of photonic components. In particular, they affect the electric field of the optical beam not only in amplitude, by splitting it into multiple optical waveguides and introducing losses, but also in phase. Memory effects, e.g., for resonant devices, will also be used to mix up incoming signals in time with previous ones, therefore having past bits interacting with present ones, similarly to what happens in reservoir computing. At the output of this class of architectures, we aim to use non-linear devices such as photodiodes (PDs) that are sensitive not only to the amplitude but also to the phase of the light field due to the coherence of the approach. The ASIC then processes the responses through transimpedance amplifiers (TIAs) and analog-to-digital converters (ADCs). The collected signals are then corrected by various means, for example, using error correction codes (ECCs) to account for potential deviations in the case of weak PUFs (see Fig.~\ref{fig:security_HW_SW}). Finally, the post-processed responses are sent to the software layer by means of a RISC-V interface. We recently proposed and demonstrated a PUF architecture based on microring resonator arrays according to the scheme of Fig.~\ref{fig:PUF_high_level} capable of achieving very good statistical performance (fractional Hamming distance close to 50\% intra and inter-device and good score for various NIST tests) \cite{Jimenez_2023_FIO}. This architecture worked at 25 Gbit/s (based on a Mach-Zehnder modulator) and was considered on a Silicon-On-Insulator (SOI) platform. Future work in NEUROPULS will further investigate these architectures and how to achieve not only good statistical properties, but also improve their robustness against fluctuations and machine learning attacks with respect to the full platform. In particular, the next section will discuss some techniques that we are considering in NEUROPULS to improve the reliability of these primitives.

\subsection{Techniques for PUF quality improvement}
Vinagrero et al. in~\cite{10224877} developed a simulation-based filtering algorithm that computes the probability of bit aliasing, exploiting the inherent relationship between reliability and bit aliasing. For RO-based PUFs, pairs of ROs with low frequency differences are known to be unreliable. However, pairs of ROs with the highest frequency difference cannot be chosen either, as the influence of process variability will be weakened, thus making the responses very similar among different devices.

Indeed, frequency differences close to the response selection boundary tend to provide the maximum amount of entropy due to the Gaussian nature of the random source, but can be deemed unreliable since they are more prone to bit flips due to noise or environmental conditions. However, frequency differences that lie further from the selection boundary could be deemed biased (aliased). Extreme values of frequency difference could be present in multiple devices because of the lower effect of process variability and, consequently, present aliasing.

\begin{figure}[h!]
    \centering
    \includegraphics[width=\linewidth]{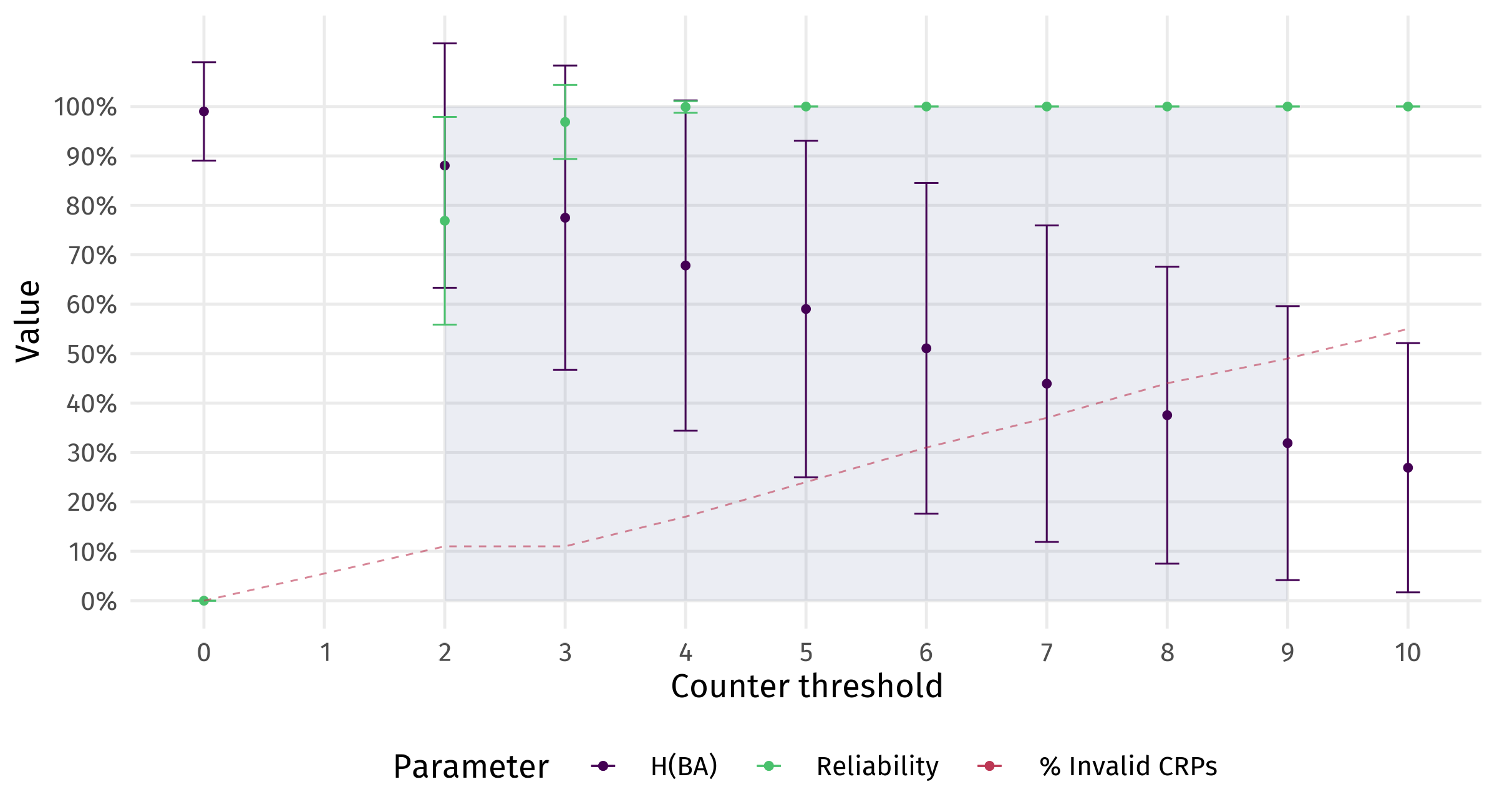}
    \caption{Relationship between bit aliasing and reliability taking
into account the counter threshold}
    \label{fig:ba_vs_rel}
\end{figure}

Fig.~\ref{fig:ba_vs_rel} reports an example of this phenomenon \cite{10224877}. Bit-aliasing is represented as Shannon entropy; values close to 1 represent no bit-aliasing, while values close to 0 represent aliasing. The shaded area represents a good trade-off between bit aliasing, reliability, and the number of challenge-response pairs (CRPs).
This technique can be tailored to different PUF architectures by adapting the threshold selection to the key generation mechanism of the PUF. For delay-based PUFs, this filtering technique is based on a threshold on the count or frequency difference of the RO pair used. In NEUROPULS, we will use a similar approach, where instead of considering a counting threshold, we will consider a threshold dependent on the amplitude of the photocurrent read at the PD. Other techniques will also be considered to reduce the effect of fluctuations, such as introducing a photonic sensor for temperature measurement and considering this additional parameter when evaluating the genuinity of the responses. Hardware approaches based on the temperature controller will also be used to reduce reliability concerns.




\section{Proposed Services and Related Security Protocols} \label{sec:services}
\subsection{Mutual authentication}
Authentication is the first step in secure communication, which consists of verifying the identity of a participant (e.g., a device) before exchanging sensitive data with the participant. The context of NEUROPULS includes two main roles: the Device to be verified and an external entity acting as the Verifier. This context imposes two requirements: First, the authentication shall be {\em mutual}, i.e., considering that sensitive data travel in both directions, the Device and the Verifier should verify each other's identity. Second, the authentication process shall be lightweight because the resources on the device are constrained.

Existing authentication strategies based on PUFs require the Verifier to store a large database of CRPs for each device, as first described in seminal works on PUFs~\cite{doi:10.1126/science.1074376,DBLP:conf/ccs/GassendCDD02} and detailed by Suh et al.~\cite{DBLP:conf/dac/SuhD07}, or to exploit heavier protocols~\cite{DBLP:conf/icoin/JungJ13,DBLP:conf/iwcmc/LeeLA13}.
However, to meet the lightweight requirement, we decided to adopt a different strategy, that is, {\em HSC-IoT}, the authentication procedure proposed by Hossain et al.~\cite{DBLP:conf/mobilecloud/HossainNH17}. Their idea is to use only a single CRP as a shared secret between the Device and the Verifier to support mutual authentication and to update it after each use with a fresh CRP.

Practically, the first CRP is shared at manufacturing time and is meant to support the first actual authentication session. At each actual authentication session, the Device uses a fresh CRP that is based on the response of the previously used CRP.
The new response is sent to the Verifier in encrypted form, and if mutual authentication succeeds, the current CRP is updated on both the Device and the Verifier.


\begin{figure}
    \centering
    \includegraphics[width=1\linewidth]{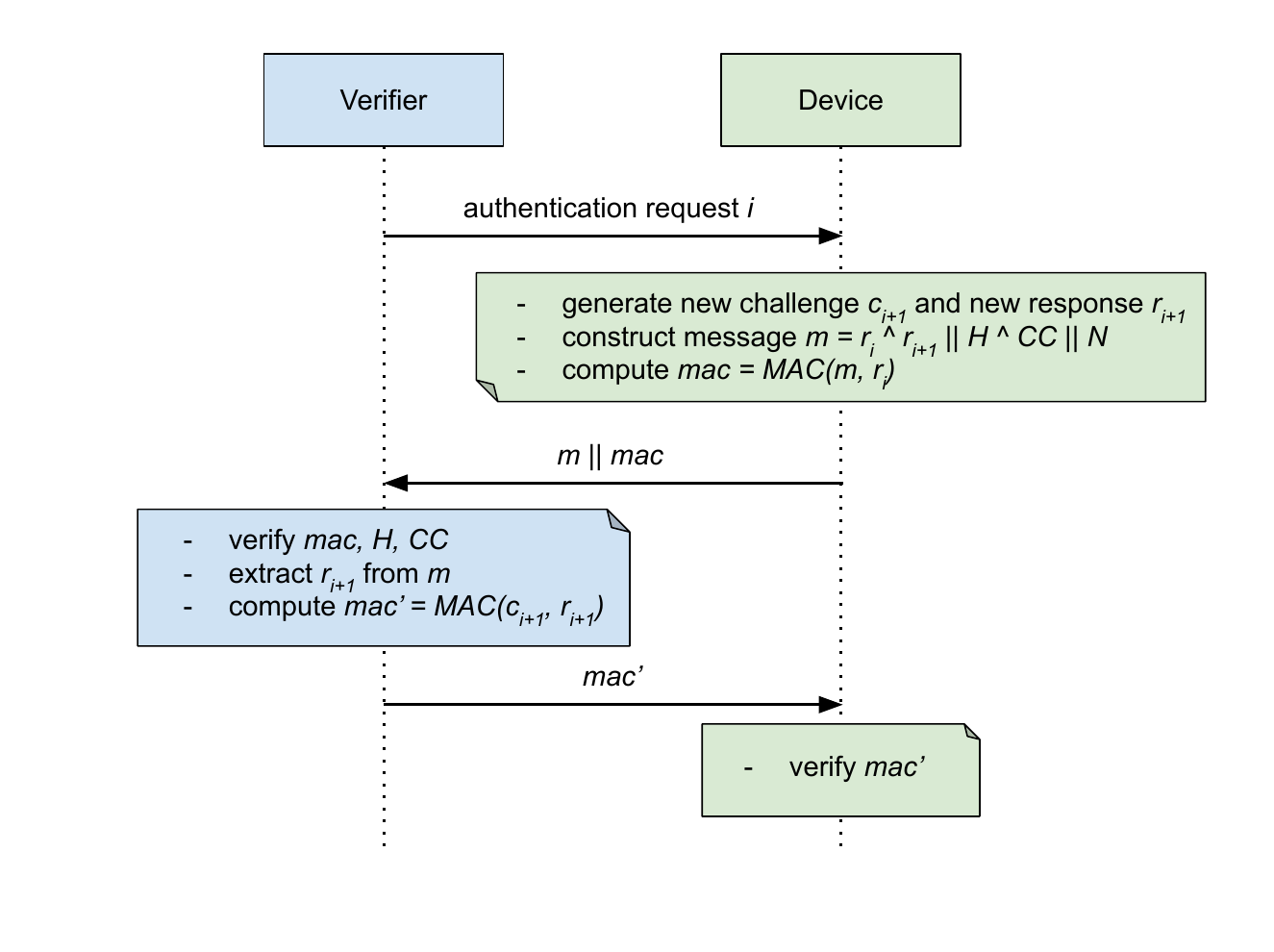}
    \caption{Session $i$ of the mutual authentication protocol.}
    \label{fig:mutual_auth}
\end{figure}
Fig.~\ref{fig:mutual_auth} contains a UML sequence diagram showing messages exchanged between the Device and the Verifier according to the mutual authentication protocol.
The protocol starts with the authentication request from the Verifier. The Device derives the new challenge $c_{i+1}$ from the current response $r_i$, using it as a seed for a pseudo-random number generation (RNG) function known to both participants, $c_{i+1} = RNG(r_i)$. The Device computes a message $m$ containing the new response $r_{i+1}$, XORed with the current response $r_i$. In the figure, the operator \textasciicircum{} denotes the XOR operation and $||$ denotes concatenation. The message may also contain proof of the integrity of the software, such as the hash of the memory $H$ and a clock count $CC$ that represents the time needed to perform a given task. The message can contain a nonce $N$ for freshness. This message $m$ is then sent to the Verifier along with its MAC signature, calculated using the $MAC(data, key)$ function, whose first argument is the data to sign and the second is the key, $r_i$ in this case. Now, the Verifier can authenticate the Device by checking the message signature using the secret $r_i$. The new response $r_{i+1}$ is derived from $m$ and stored in the Verifier to be used as a shared secret in the next authentication session.
Finally, the Verifier authenticates to the Device by demonstrating that it knows the new secret $r_{i+1}$, which is used to sign $c_{i+1}$ (generated using $r_{i+1}$) through the MAC function again.

This protocol only needs one CRP to be known by the Verifier at any point, which is more scalable than other solutions that require a large database of CRPs. In addition, CRPs are kept confidential because they are never exchanged in clear text. Although this protocol is very lightweight, it is designed to resist many attacks, as discussed by Hossain et al.~\cite{DBLP:conf/mobilecloud/HossainNH17}.

\subsection{Software attestation}
Remote software attestation validates the integrity status of remote computing devices without relying on secure hardware components such as Trusted Platform Modules (TPMs). 
Such specialized components are often unsuitable for devices with limited resources \cite{5727971,aman2020hatt}. 
This approach enables remote systems to detect malicious or unintended changes in the firmware, software, or hardware that operates on these devices.

Attestation mechanisms generally send a hash of the device’s memory to the Verifier to prove that the device is not compromised \cite{feng2018aaot}.
An example of this is given in the previous section; during mutual authentication, the algorithm can include a hash of the memory, which provides some proof of the device's integrity.
In this section, we describe a more powerful approach that, however, imposes stronger assumptions, i.e., it leverages an ideally reliable strong PUF and on a PUF model available to the Verifier. To avoid attacks where a device hides its compromised memory regions from verification (e.g., by moving these regions around while the algorithm hashes the uncompromised memory), attestation protocols often employ temporal constraints that guarantee the unfeasibility of these attacks \cite{aman2018att}. An important part of these protocols is the root of trust, a part of the system proven not to be compromised, on which the protocol can base its operations.
Without being able to use secure hardware modules (i.e. TPMs) as a root of trust, many modern protocols have started adopting PUFs as an alternative.
In our framework, we leverage the photonic PUF (pPUF) embedded in the neuromorphic accelerator to generate a large number of CRPs that can then be used to hash different areas of the device’s memory.



The Verifier starts the attestation by crafting a message, the attestation request, that contains a timestamp $t$ and a challenge $c_1$. 
The message is then sent to the Device, which then promptly starts the attestation process by using the issued challenge to compute a response $r_1$ in the pPUF.
The response is combined with the timestamp as the seed for an RNG that generates the random walk in memory: $m_1, \ldots, m_n = RNG(r_1 + t)$.
A secure hash algorithm is then used with the initial chunk of memory $m_1$ on the device and $r_1$, generating the first hash $h_1 = HASH(m_1, r_1)$, while $r_1$ is used simultaneously as the next challenge for pPUF, as $r_1 = pPUF(r_1)$.
All subsequent hashes depend on the previously calculated ones: $h_{i+1} = HASH(m_{i+1}, r_{i+1}, h_i)$.
The inherent speed of the pPUF (at least 5 Gb/s) guarantees that the constant challenge-and-response generation never slows down the protocol, so the temporal constraints of our approach can be stricter than those found in previous work.
After exhausting all memory regions, the final hash, $h_n$, is sent to the verifier.
This protocol minimizes the network load by having a very small footprint in both the attestation initiation and its finalization, which allows our temporal constraints to be mostly focused on the speed of the iterative hash function.
The Verifier has a copy of the uncompromised device memory and a model of the pPUF, so it can start to calculate $h_n$ right after generating the attestation request.
After receiving $h_n$ from the Device, the Verifier checks that its value is correct and that the attestation did not exceed its temporal constraints; if both of these requirements are satisfied, the attestation is successful. Otherwise, a new request is issued and the protocol restarts with a new timestamp and challenge.
\subsection{Neural network configuration and data encryption}
Another important security requirement is the confidentiality of the actual data processed by the accelerator and the confidentiality of the neural network configuration. To meet this confidentiality requirement, encryption encodes the data in all communications with external parties and all the software running on the device.
The NN configuration, the input data to the device, and the computation output are encrypted using the secret keys described in Section~\ref{sec:security_primitives}. This key is never exposed to the software layer, but encryption and decryption occur on the hardware in the implementation of the security primitives shown in Table~\ref{tab:nn_encryption}.
\texttt{load\_network} receives the neural network configuration in encrypted form. The configuration is decrypted in hardware and loaded in the accelerator.
\texttt{execute\_network} takes the input as a decrypted parameter and fed to the accelerator. Then, the computation result is encrypted and returned by this function.
\begin{table}
    \centering
    \begin{tabular}{lll}
        Function name & Parameters & Results \\
        \hline
        \scriptsize\texttt{load\_network} & \scriptsize\texttt{ciphered\_network} & \\
        \scriptsize\texttt{execute\_network} & \scriptsize\texttt{ciphered\_input} & \scriptsize\texttt{ciphered\_output}\\\\
    \end{tabular}
    \caption{Functions to load and execute a neural network}
    \label{tab:nn_encryption}
    \vspace{-0.9cm}
\end{table}
As specified by the function signatures, data are never exposed in plaintext to the software. Data are decrypted internally by the device using primitives that never leave plaintext in the memory after execution, thus preserving confidentiality even against an internal attacker capable of reading the RAM.
\section{Strengths Against Attacks} \label{sec:attacks}
The core of the security services to be provided in NEUROPULS are supported by the use of PUF intrinsically bound at both the PIC and the ASIC levels. This protects our NN accelerator from tampering attacks where one malicious chip could replace the genuine PIC or control ASIC. The robustness of this security mechanism is also possible thanks to the effectiveness achieving by the physical connection between chips that are highly dependent on the packaging (e.g., using high-frequency wire-bonding) as well as components such as transimpedance amplifiers (TIAs) and ADC modules that further modify the photonic PUF response, it is possible to generate a composite response from the 2 chips, which can be used to assess the genuine character of the accelerator as a whole.

In NEUROPULS, we will investigate the robustness of the security layers against various types of attacks, with a particular focus to attacks targeting the PUF, ranging from machine learning modeling to side-channel attacks. 
We will specifically look at attacks that tamper with the responses received on the PD array or at how laser power levels can be altered to produce responses that can provide insights into the inner working mechanisms of the PUFs.
In particular, effects such as bit flips will be addressed from both a theoretical and an experimental point of view in the strings sent from the driving ASIC to the PIC. It is important to note that integrated photonics have a striking advantage compared to electronic technologies. Photonic PUFs can transfer information while being manipulated and do not suffer from RF leakage, contrary to many electronic PUF solutions. In the latter case, RF signals can be detected, for example, from the Si substrate, as was pointed out in various works. Therefore, by performing a power analysis, it was possible to extract key information about PUF behavior and thus carry out modeling attacks \cite{shamsoshoara2020survey,ruhrmair2014efficient}. The capability of transferring information in photonic waveguides where signals leak out only a few hundred nanometers hinders side-channel attacks. Although these attacks can still potentially occur at the interface between the PIC and the ASIC (see Fig.~\ref{fig:PUF_high_level}), the analogous and high-speed nature of such signals and the fact that the signals will be processed further at the ASIC level brings several additional layers of complexity that require a very expensive way to break such solutions. Furthermore, although in our prototype the ASIC and the PIC will be connected by wire bonding, approaches in which the electronics are integrated with the photonics in a monolithic way will render this potential attack point useless as the information will be electronically manipulated locally where it was processed optically \cite{atabaki2018integrating,sun2015single}.
Furthermore, the photonic PUF discussed in Fig.~\ref{fig:PUF_high_level} operates in a time domain; therefore, its response is present only during the interrogation time and then disappears. Therefore, attacks based on the remanence decay time cannot be used, as in SRAM PUFs that share memory with other functionalities \cite{zeitouni2015remanence}. Another strength comes from the fact that the response is present in the PUF for a very short period of time (below 100 ns), making its potential extraction very complex.

Machine learning modeling of PUFs is another common way to attack PUFs. In such a case, by acquiring a sufficiently large number of CRPs (for strong PUFs), the adversary can build a model to predict the response to the next challenge. This approach could then be used to impersonate the system containing the PUF, thus breaking the security of the layer and the system as a whole. These attacks have been particularly successful against common types of PUF, such as PUFs with ring oscillators (ROs) or arbitrators \cite{ruhrmair2010modeling}. The main weakness of this type of PUF lies in the relatively small number of components and variables that participate in processing the challenge to generate the response. Photonic PUFs are expected to provide a greater gain with respect to modeling attacks because of the much larger number of components and, especially, variables that participate. Various examples of optical/photonic PUFs resistant to this type of attack have been previously shown in the literature \cite{bosworth_unclonable_2020}. In NEUROPULS, we will investigate strong PUFs that exploit non-linearities at different levels as well as architectural solutions that rely on the combination of a strong and a weak PUF to encrypt the challenges before entering the photonic PUF as we previously proposed for purely electronic PUFs \cite{vatajelu2019encryption}. To further enhance the robustness and security of the PUF use in the authentication process, while also generating session keys for the data encryption, an Authentication and Key Agreement (AKA) protocol can also be considered. AKA can protect the PUF responses in such a way that an attacker cannot guess or brute-force the protocol to find the Challenge-Response Pair. One approach to achieve this is to see the CRP as a low-entropy shared secret. With this, we can consider the use of the well-established and secure EKE protocol to achieve both mutual authentication and key exchange, to be used in the secure channel implementation.
This approach protects against most possible attacks to the CRP while providing perfect forward security to the key established for data encryption. Note that this approach is computationally more expensive. However, lighter AKA protocols can be considered~\cite{AKA_survey} depending on the target security and robustness of the achieved PUF.
\section{System-Level Modeling and Benchmarking} \label{sec:benchmarking}
Building a simulator capable of modeling the behavior of security primitives, such as PUFs, requires modeling all system components (CPU, memory, accelerators) and implementing suitable benchmarks to evaluate the performance and security characteristics. Regarding system-level modeling, defining the PUF architecture and specifying its fundamental components, such as challenge-response pairs, arbiter circuits, and memory elements, together with the interface for programming them, is essential. The general structure, the number of stages and the types of components should be carefully considered. Environmental factors, including temperature, voltage, and variations in the manufacturing process, must also be simulated to account for their potential impact on the behavior of PUF. Furthermore, noise and other sources of variability should be modeled to fully assess PUF responses.

The simulator should also account for the interactions between the PUF and other system components, such as cryptographic modules and key management systems. Parameters for configuration and tuning should be included to simulate the effects of different settings on PUF performance and security. The gem5~\cite{gem5-2} simulation environment allows one to define a peripheral module connected to the RISC-V microprocessor, providing the essential infrastructure for the delivery of the programming API. Furthermore, the simulator should incorporate models for potential attacks on the PUF, as described in Section~\ref{sec:attacks}. A holistic approach to modeling and simulating a heterogeneous system is required, including RISC-V (or other ISA) CPUs and electronic or photonic accelerators~\cite{risc-v-summit-chatzop,10173974}. 

As defined in Section~\ref{sec:security_primitives}, reliability metrics are crucial for benchmarking, including evaluating the reliability of PUF responses under different conditions, such as environmental variations and the effects of aging. The gem5-provided log facility allows data collection to assess entropy, uniqueness, and response uniformity to ensure that the modeled PUF provides a likelihood of random and unique identifiers. Additionally, error rates, including false positive and false negative rates, should be analyzed to gauge the PUF's reliability. Additionally, throughput, latency, and power consumption measurements are essential to understand the practical performance of PUFs in real-world applications. Lastly, the simulator should evaluate the PUF's robustness against various attacks, encompassing machine learning-based, side-channel, and invasive attacks. By incorporating these considerations into the simulator, a comprehensive tool can be created to analyze and optimize the performance and security characteristics of PUFs in various scenarios.



\bibliographystyle{IEEEtran}
\bibliography{security}

\end{document}